# Optimal Separable Algorithms to Compute the Reverse Euclidean Distance Transformation and Discrete Medial Axis in Arbitrary Dimension

David Coeurjolly and Annick Montanvert

**Abstract**—In binary images, the Distance Transformation (DT) and the geometrical skeleton extraction are classic tools for shape analysis. In this paper, we present time optimal algorithms to solve the reverse Euclidean distance transformation and the reversible medial axis extraction problems for $d$-dimensional images. We also present a $d$-dimensional medial axis filtering process that allows us to control the quality of the reconstructed shape.

✦

## 1 INTRODUCTION

In binary images, the distance transformation (DT) and the geometrical skeleton extraction are classic tools for shape analysis [1], [2]. The distance transformation consists of labeling each pixel of an object with the distance to the closest pixel of its complement (also called the *background*). Obviously, a distance transformation algorithm is deeply linked to the underlying metric and the objective when we define a new metric or approximate the Euclidean distance is to achieve isotropic behavior of the DT while preserving efficient algorithms to compute the DT.

In the digital image literature, trade-offs were considered between computation time and the quality of the Euclidean distance approximation.

Hence, for the DT problem, we can consider distances based on chamfer masks [2], [3], [4], [5] or sequences of chamfer distances [1], [6], [7], the vector displacement-based Euclidean distance [8], [9], [10], [11], the Voronoi diagram-based Euclidean distance [12], [13], [14], [15], or the square of the Euclidean distance [16], [17], [18]. From a computational point of view, several of these methods lead to time optimal algorithms to compute the error-free Euclidean Distance Transformation (EDT) for $d$-dimensional binary images [12], [14], [17], [18], [15]: The extension of these algorithms is straightforward since they use separable techniques to compute the DT; $d$ one-dimensional operations—one per direction of the coordinate axis—are performed.

The skeleton and the medial axis are usual and convenient representations for shape description or recognition purposes [19], [20]. In continuous space, several equivalent definitions of the skeleton exist: We can consider the *prairie*

*fire model* with a wavefront propagation initiated at the shape boundary; then skeleton points are the locations of the "self-intersections" of the wavefront. In this case, a classification of skeleton points can be obtained while identifying intersection and transition cases [21], [22]. Another approach is based on the detection of ridges and peaks on the distance map surface. A third model defines the skeleton as the set of center pixels of maximal disks covering the shapes: A maximal disk is a disk contained in the shape not entirely covered by another disk contained in the shape.

From these definitions in continuous space came different categories of methods in discrete space. From the detection of symmetries came the Voronoi diagram approaches in computational geometry [23]; from the detection of ridges and peaks came some variational approaches [24], [25], [26], [27]; from the prairie fire model came the iterative pealing model providing binary skeletons (see [28] for a complete bibliography on the subject). Finally, in the digital plane, we have methods based on the discrete DT to extract the medial axis (MA for short). Indeed, given a binary shape, the DT value at a point $p$ corresponds to the radius of the largest ball centered at $p$ contained in the shape. Many discrete implementations of this approach have been proposed either for chamfer distances [1], [3], [29], [30] or for the Euclidean distance [31], [32], [33], [34]. In digital space, the MA is a convenient tool to represent shapes since it is reversible: From the MA points, we can exactly reconstruct the original shape.

In this paper, we focus on the study and new results in this last category of methods. More precisely, we investigate the $d$-dimensional medial axis extraction upon the error-free Euclidean distance. An important problem related to the MA extraction is the Reverse Euclidean Distance Transformation (REDT). Furthermore, the resolution of these two problems will be linked by our approach. Formally, given a set of points associated with their Euclidean distance values, how can we efficiently reconstruct the shape resulting from the overlapping of the corresponding balls? A time optimal algorithm is proposed to solve this problem. Based on this process, we also present a time optimal algorithm to compute a subset of the medial axis on $d$-dimensional shapes: Even if the MA is

----

- *D. Coeurjolly is with the Laboratoire LIRIS UMR-CNRS 5205, Université Claude Bernard Lyon 1, 43 Bd du 11 novembre 1918, F-69622 Villeurbanne, France. E-mail: david.coeurjolly@liris.cnrs.fr.*
- *A. Montanvert is with the Laboratoire LIS UMR-CNRS 5083, INPG 61 rue de la Houille Blanche, BP 46, F-38402 St. Martin d'Hères, France. E-mail: annick.montanvert@lis.inpg.fr.*

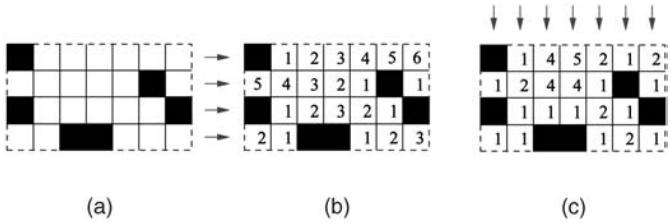

Fig. 1. Illustration of the SDT algorithm: (a) the binary image $P$, (b) the map $G$ resulting from Step 1 (absolute EDT), and (c) the final SDT map $H$ after the last process along the $y$-axis.

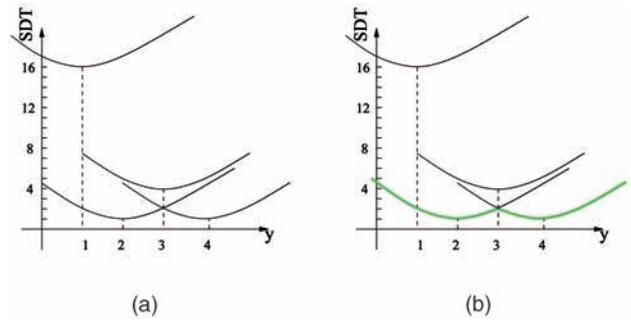

Fig. 2. Illustration of the computation of the mixing step as a lower envelope extraction: Let $[4, 1, 2, 1]$ be a column of $G$ after Step 1 (fifth column in Fig. 1), (a) the set of parabolas $g(i, y)^2 + (j - y)^2$ and (b) the bold curve is the lower envelope. Thus, the result of the minimization process is $[2, 1, 2, 1]$.

a reversible representation of the shape, it may not be the minimal set of disks necessary to reconstruct the shape [35], [36] (a set is minimal if it contains the minimal number of disks). We thus obtain a more compact representation with disks of a discrete object than the classic MA. To achieve generalization in higher dimensions, we investigate separable techniques to solve the REDT or the MA extraction problems.

In some applications, due to the sensitivity of the MA to small changes in the object boundary, a nonreversible but simplified description of binary objects may be of interest. In that case, a simplification procedure can be used as a postprocess [37], [23], [38]. Based on a discussion about the REDT, the MA extraction, and classic tools in computational geometry, we present a simple filtering process to simplify $d$-dimensional MA.

In Section 2, we first evoke some algorithms to solve the EDT problem for $d$-dimension in a linear time. Based on these techniques, we optimize the REDT algorithm proposed by Saito and Toriwaki [33] to obtain a true linear algorithm in Section 3. Then, in Section 4, based on an analysis of the literature, we present a time optimal algorithm that extracts a reversible subset of the classic medial axis. Finally, in Section 5, we discuss the links between these algorithms and classic tools in computational geometry, illustrated by an MA filtering procedure.

## 2 $d$-DIMENSIONAL EUCLIDEAN DISTANCE TRANSFORMATION

We first detail separable techniques to compute the Euclidean distance transformation of $d$-dimensional images.

In the 2D case: We consider a two-dimensional binary image $P$ of size $n \times n$; $\bar{P}$ denotes the complementary of $P$, i.e., the set of background pixels. The output of the algorithm is a 2D image $H = \{h(i, j)\}$ storing the squared distance transformation. For each point $(i, j)$ of the image, the squared distance transformation is given by:

$$h(i, j) = \min\{(i - x)^2 + (j - y)^2; 0 \leq x, y < n \\ \text{and } (x, y) \in \bar{P}\}. \quad (1)$$

This formulation of the problem leads to an efficient two-pass process for the squared distance transformation (SDT for short) labeling in 2D (see Fig. 1):

1. Build, from the source image $P$, a one-dimensional EDT according to the first dimension ($x$-axis) denoted by $G = \{g(i, j)\}$, where, for a given row $j$:

$$g(i, j) = \min_x\{|i - x|; 0 \leq x < n \text{ and } (x, j) \in \bar{P}\}. \quad (2)$$

2. Then, construct the image $H = \{h(i, j)\}$ with a $y$-axis process:

$$h(i, j) = \min_y\{g(i, y)^2 + (j - y)^2; 0 \leq y < n\}. \quad (3)$$

This formulation of the SDT provides a direct implementation of the $d$-dimensional SDT algorithm: We only have to compute a one-dimensional EDT for the initialization step (Step 1 of the previous algorithm) and then add, for each greater dimension, a mixing process (Step 2) that merges results of the lower dimensions. From a computational cost point of view and given a $d$-dimensional binary shape of size $n^d$, the first step can be done in linear time in the number of grid points, i.e., $O(n^d)$. For the second step, the *min* operation corresponds to a lower envelope computation of a set of parabolas. More precisely, let us suppose that we have computed Step 1 of the algorithm ($x$-axis SDT) and let $\{g(i, y)\}(0 \leq y < n)$ be a column of $G$. If we consider the set of parabolas $\mathcal{F}_y^i(j) = g(i, y)^2 + (j - y)^2$, the column $\{h(i, j)\}$ after Step 2 is exactly the lower envelope of $\{\mathcal{F}_y^i\}$ with $0 \leq y < n$ (see Fig. 2).

In [16], the authors present an $O(Avg.n^d)$ algorithm that computes each mixing step where $Avg$ denotes the average of the Euclidean distance values in the image ($Avg = O(n)$ without any assumptions based on the input image). In [17] and [18], Hirata and Meijster et al. independently present optimal algorithms to solve the *min* operation (Step 2) and thus propose a time optimal algorithm for the SDT. The authors present an $O(n)$ algorithm to compute such a lower envelope using a parabola elimination process. Finally, for a $d$-dimensional image, the dimensional mixing processes are computed in $O(n^d)$ and, thus, the global cost to compute the SDT based on this approach is $O(n^d)$. We recall, in Fig. 3, the optimal Meijster et al.'s algorithm in dimension 2. To detail the notations, $\mathcal{F}_y^i(j)$ denotes the parabola $g(i, y)^2 + (j - y)^2$ (simply denoted $\mathcal{F}_y(j)$ when the context fixes the parameter $i$). The function $\mathcal{S}ep^i(u, v)$ (or simply $\mathcal{S}ep(u, v)$) represents the coordinate of the intersection point between two parabolas. Hence, according to [18],

$$\mathcal{S}ep(u, v) = (v^2 - u^2 + g(i, v)^2 - g(i, u)^2) \text{ div } (2(v - u)). \quad (4)$$

For the second step of this algorithm, we use a stack (denoted $s[q]$) to store the indexes of the parabolas on the lower envelope. When we scan the rows for a given column $i$, a new parabola may invalidate some parabolas in the lower envelope stack (while loop in lines 4-6) and may be inserted into $s[q]$ (line 12). To complete the SDT computation, we perform a final scan of the lower envelope parabolas to set the correct values to the distance map $h(i, j)$ (lines 16-21).

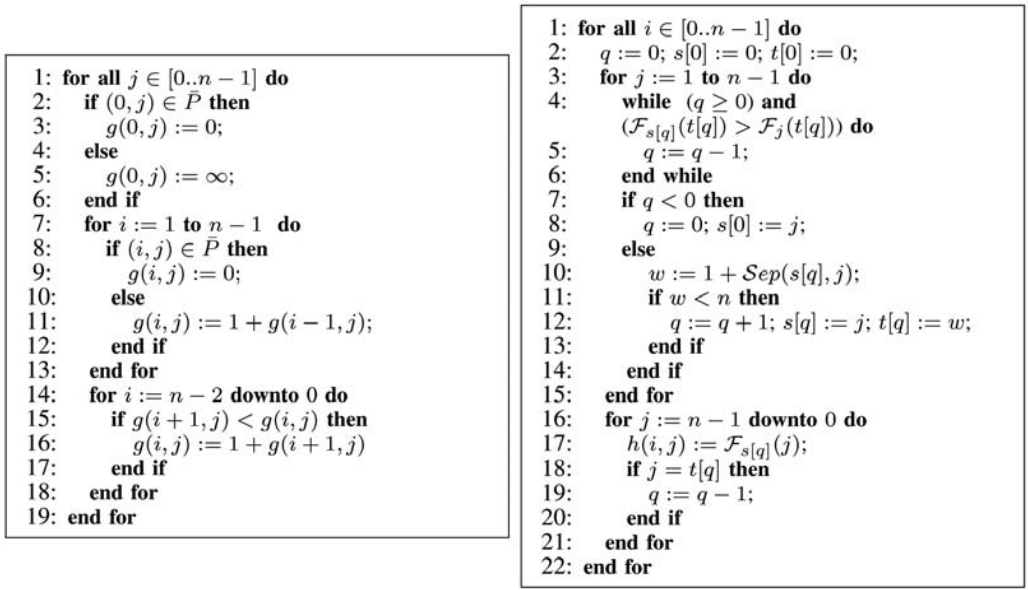

Fig. 3. Pseudocode of the optimal SDT: (a) Step 1 according the the $x$-axis and (b) Step 2 according to the $y$-axis.

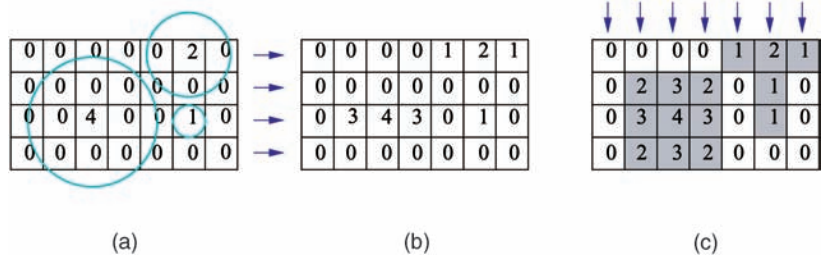

Fig. 4. Illustration of the REDT algorithm: (a) the image $F$, where we only consider the set of disks, (b) the map $G$ after the first step along the $x$-axis, and (c) the final reconstruction $H$.

## 3 REVERSE EUCLIDEAN DISTANCE TRANSFORMATION

### 3.1 Definitions

Let us consider $L$ as a set of $l$ points $\{(x_m, y_m)\}_{1 \le m \le l}$ and $r(x_m, y_m)$ the squared Euclidean distance value associated with the pixel $(x_m, y_m)$. In other words, a point $(i, j)$ belongs to $P$ if it belongs to at least one disk whose center is a point $m$ of $L$, with radius $\sqrt{r(x_m, y_m)}$. Hence, the REDT of $L$ consists of obtaining the set of points $P$ such that

$$P = \{(i, j) \mid (i - x)^2 + (j - y)^2 < r(x, y), (x, y) \in L\}. \quad (5)$$

The question is to efficiently compute the REDT. Let $F = \{f(i, j)\}$ be a picture of size $n \times n$ such that $f(i, j)$ is set to $r(i, j)$ if $(i, j)$ belongs to $L$ and 0 otherwise. In [33], the authors show that (5) is equivalent to

$$P = \{(i, j) \mid \max\{f(x, y) - (i - x)^2 - (j - y)^2\} > 0; \\ 0 \le x, y < n \text{ and } (x, y) \in F\}. \quad (6)$$

Hence, if we compute the map $H = \{h(i, j)\}$ such that

$$h(i, j) = \max\{f(x, y) - (i - x)^2 - (j - y)^2; \\ 0 \le x, y < n \text{ and } (x, y) \in F\}, \quad (7)$$

we obtain $P$ by extracting from $H$ all pixels of strictly positive values. So, to build $H$ from $F$, we can decompose the computation into two one-dimensional steps (see Fig. 4):

1. Build from the image $F$ the picture $G = \{g(i, j)\}$ such that

$$g(i, j) = \max_x \{f(x, j) - (i - x)^2, 0 \le x < n\}. \quad (8)$$

2. Build from $G$ the picture $H$ such that

$$h(i, j) = \max_y \{g(i, y) - (j - y)^2, 0 \le y < n\}. \quad (9)$$

To prove this decomposition, we have to substitute (8) into (9) and we obtain (7). Note that this process can easily be extended to $d$-dimensional images, we just have to compute $d$ one-dimensional maximization steps. An illustration of the overall algorithm for $d = 2$ is given in Fig. 4.

In [33], Saito and Toriwaki use their algorithm presented in [16] to compute the REDT and they obtain a computational cost in $O(Avg.n^d)$ for a $d$-dimensional image. Indeed, if we change the minimization process in (3) to a maximization procedure as in (8) or (9), the SDT algorithm can also be used to solve the REDT problem.

In the next section, we detail a new algorithm whose complexity is $O(n^d)$ to compute the REDT.

## 3.2 Optimal REDT Algorithm

The basic idea of the optimal REDT algorithm is to use the parabola elimination process described in Section 2 to compute maximization steps. We detail the optimization of Step 1 defined by (8) all other steps can be easily deduced. First of all, for a given column $j$ of $F$, we define the function describing a parabola as:

$$\mathcal{F}_x^j(i) = f(x, j) - (i - x)^2. \qquad (10)$$

We simply use the notation $\mathcal{F}_x(i)$ when the context fixes the parameter $j$. We also need the function that computes the abscissa of the intersection between two parabolas. Thus, point $i$ such that $\mathcal{F}_u(i) \geq \mathcal{F}_v(i)$ with $u < v$ is given by:

$$\mathcal{S}ep(u, v) = (u^2 - v^2 - f(u, j) + f(v, j)) \text{ div } (2(u - v)). \qquad (11)$$

Based on these elementary functions, the algorithm presented in Fig. 5 computes the upper envelope of the parabolas $\{\mathcal{F}_x\}$.

This algorithm is derived from the one presented in Fig. 3b to compute the SDT: The array $s$ contains the set of parabola apexes of the upper envelope and $t$ the intersection abscissa between two consecutive parabolas in $s$. In lines 3–15, we compute the upper envelope and those arrays $s$ and $t$ and, in lines 16–21, we construct the map $G$ using $s$ and $t$. The computational cost of this upper envelope extraction is $O(n)$ if $n$ is the size of a row in $F$. Finally, we can use this algorithm to compute (9) and construct $P$ by thresholding $H$ in $O(n^2)$ if $F$ is an $n \times n$ image. More generally, if we apply it for all one-dimensional maximization steps, we have a global complexity in $O(n^d)$ for a $d$-dimensional image, which is time optimal.

This formulation of the REDT problem as a parabola upper envelope computation will be also used in the next section to extract the discrete medial axis.

## 4 EUCLIDEAN MEDIAL AXIS EXTRACTION

### 4.1 Definitions and State-of-the-Art

To characterize the medial axis as defined in the literature (see, for example, [2]), we need the following definitions:

**Definition 1 (Maximal ball).** *A maximal ball is a ball contained in the shape not entirely covered by another ball contained in the shape.*

Based on this property, the medial axis is defined by:

**Definition 2 (Medial axis).** *The medial axis of a shape is the set of maximal ball centers contained in the shape.*

It is obvious from this definition that MA is reversible. Given a binary shape, the SDT value at a point $p$ corresponds to the square of the radius of the largest ball centered at $p$ contained in the shape. Hence, the MA is a subset of the set of balls defined by the DT. The main bottleneck is thus the ball inclusion test: Given a ball in the DT, we have to decide if the ball is covered by another one or not. Of course, the MA is dependent on the space—discrete or continuous—and on the metric that is used. We continue in this section by the analysis of the discrete medial axis based on the Euclidean metric.

In [32], the authors first present a test in dimension 2 to decide if a disk $A$ covers another disk $B$. Then, they use this inclusion test on each couple of disks defined in the SDT to extract the MA. Even if the authors provide optimization to reduce the number of inclusion tests, the computational cost is high: In the worst case, the computational cost of the inclusion test is proportional to the radii of the disks and we have $N^2$ inclusion tests if $N$ is the number of points in the SDT.

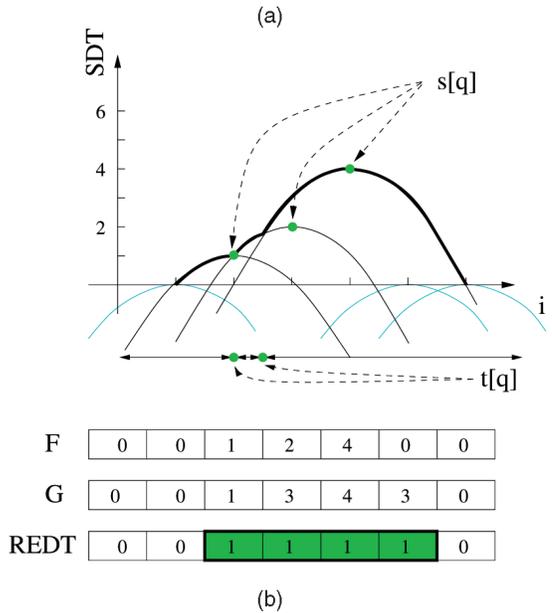

(a)

(b)

Fig. 5. Pseudocode of the (a) optimal one-dimensional upper envelope parabola computation and (b) an illustration of the notations and an example in one dimension.

In [31], [34], the authors use a lookup table to implement the inclusion test. This table is indexed by the radius of the disk $B$ with center $p$ and a direction $\vec{v}$ and contains the minimum radius of the disk $A$ with center $p + \vec{v}$ such that $A$ covers $B$. This lookup table is widely used to extract the chamfer metric-based medial axis [2], [3], [4]. In that case, the set of directions is finite and given by the mask directions. Using the Euclidean metric, if we suppose that values of the SDT are less than a given number $R_{max}^2$, the authors of [34] present an algorithm to compute the lookup table in dimensions 2 and 3. The main drawbacks of this method are that the entire precomputed table must be loaded into the memory and that the size of this table considerably increases with the number $R_{max}$ and the dimension. Indeed, the Euclidean metric makes the set of directions $\{\vec{v}\}$ unbounded: No local test exists to detect if a disk is maximal based on the SDT.

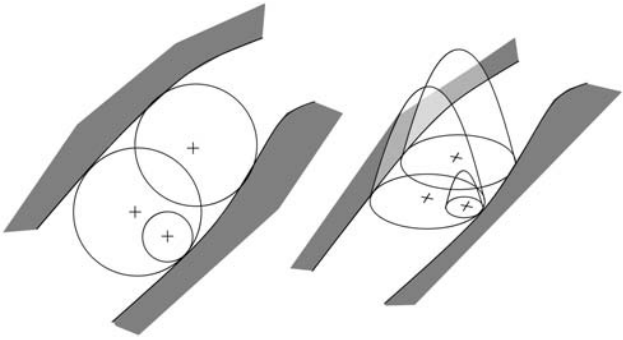

Fig. 6. Illustration of the equivalence in the continuous plane between maximal disks and maximal elliptic paraboloids.

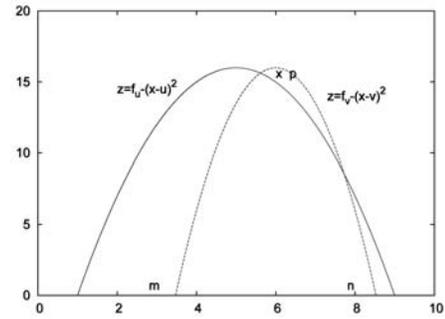

Fig. 7. Notations for the proof of Proposition 1.

If the SDT values are greater than $R_{max}^2$, the MA may contains nonmaximal balls.

In the next sections, we present three optimal separable techniques to extract the MA whatever the dimension. Instead of considering local inclusion tests, the proposed algorithm performs a global extraction of maximal balls.

## 4.2 Medial Axis and Upper Envelope of Elliptic Paraboloids in the Continuous Space

In [33], Saito and Toriwaki define what they called a geometrical Euclidean skeleton based on elliptic paraboloids in dimension 2. Such an elliptic paraboloid of center $(i, j)$ and height $q(i, j)$ is given by the following equation:

$$0 \le z < q(i, j) - (x - i)^2 - (y - j)^2. \tag{12}$$

The intersection between such a domain and the plane $z = 0$ is a disk of center $(i, j)$ and radius $\sqrt{q(i, j)}$. We say that an elliptic paraboloid is contained in a shape $S$ if the disk of center $(i, j)$ and radius $\sqrt{q(i, j)}$ is contained in $S$. Note that the authors use the term *skeleton* to describe a geometric object when, usually, *skeleton* means topology preserving representation.

Let $Q = \{q(i, j)\}$ be a SDT of the shape. The Saito and Toriwaki's skeleton, denoted $Sk$, is defined by

$$\begin{aligned} Sk = \{(i, j) \mid \exists (x, y), (i - x)^2 + (j - y)^2 < q(i, j), \\ \text{and } \max_{(u, v)} \{q(u, v) - (x - u)^2 - (y - v)^2\} \\ = q(i, j) - (x - i)^2 - (y - j)^2\}. \end{aligned} \tag{13}$$

In other words, $Sk$ is the set of elliptic paraboloids that belong to the upper envelope (in dimension 2) of all of the elliptic paraboloids whose heights are given by the squared distance transformation.

In the following, we first prove that Saito and Toriwaki's skeleton is a subset of the medial axis in the continuous plane (see Fig. 6).

**Definition 3 (Maximal elliptic paraboloid).** *A maximal elliptic paraboloid is an elliptic paraboloid contained in the shape not entirely covered by another elliptic paraboloid contained in the shape.*

Note that this object can be generalized to $d$-dimension shapes.

**Proposition 1.** *Let $(i, j)$ be a point in a continuous shape and $q(i, j)$ be a number. The disk $D$ of center $(i, j)$ and radius $\sqrt{q(i, j)}$ is maximal if and only if the elliptic paraboloid $P$ of center $(i, j)$ and height $q(i, j)$ is maximal.*

**Proof.** Note that $D$ is the intersection between $P$ and $z = 0$. We first prove the left to right implication. If we suppose that $P$ is not maximal, there exists another elliptic paraboloid $P'$ such that $P'$ contains $P$. Thus, the intersection $D'$ between $P'$ and the plane $z = 0$ contains the intersection $D$ between $P$ and the same plane. Hence, there exists a disk $D''$ contained in the shape that contains $D$ and, so, $D$ is not maximal.

Conversely, we suppose that $D$ is not maximal. Hence, there exists a disk $D''$ such that $D''$ contains $D$. We denote by $P''$ the elliptic paraboloid, uniquely defined by $D''$, such that $D''$ is the intersection between $P''$ and $z = 0$. If we suppose that $P''$ does not contain $P$, there exists a point $p \in P$ such that $p \notin P''$. Let us consider the intersections between $P$ and $P''$ in the plane $H$ perpendicular to $z = 0$ that contains $p$ and the center of $P''$. In the plane $H$ and using the elliptic paraboloid definition, $P$ (respectively, $P''$) leads to the domain

$$0 \le z < f(u) - (x - u)^2 \left( \text{resp. } 0 \le z < f(v) - (x - v)^2 \right), \tag{14}$$

with $u, v, f(u), f(v) \in \mathbb{R}$ (see Fig. 7). Since $H$ contains the center of $P''$ and $p$, these domains are not empty. Using the notations of Fig. 7, $D''$ contains $D$ implies that both $m$ and $n$ belong to $D''$. Furthermore, since $p$ does not belong to $P''$, the two parabolas given by (14) must have two intersection points $a$ and $b$. However, using (14), such parabolas only have one intersection point if $u \ne v$. Since the upper parts of the parabolas are excluded, $u = v$ implies that the intersection is empty. Hence, such a point $p$ does not exist and, thus, $P''$ contains $P$, which finally proves that $P$ is not maximal. Note that this proof can be generalized to other dimensions since we have transformed the problem into a one dimension parabola intersection. □

Hence, in the continuous plane, maximal balls and maximal elliptic paraboloids coincide. Using Proposition 1, we can deduce the following corollary:

**Corollary 1.** *In the continuous plane, $Sk$ is a subset of the medial axis. Furthermore, the original figure can be reconstructed by $Sk$.*

**Proof.** First of all, all elliptic paraboloids that belong to the upper envelope are maximal by the definition of such an

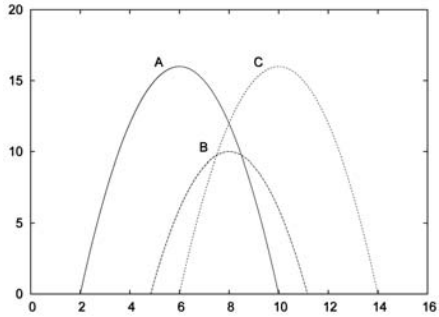

Fig. 8. Illustration for Corollary 1: The maximal ball defined by $B$ does not belong to $Sk$.

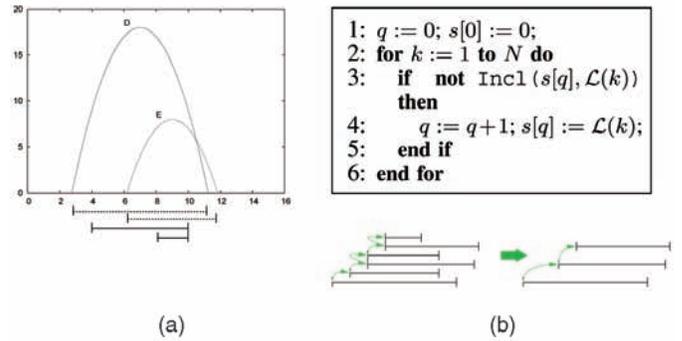

```
1: q := 0; s[0] := 0;
2: for k := 1 to N do
3:    if  not Incl (s[q], 𝓛(k))
      then
4:       q := q+1; s[q] := 𝓛(k);
5:    end if
6: end for
```

(a)          (b)

Fig. 9. An illustration of the difference between the skeleton $Sk$ and the discrete medial axis. (a) Dashed segments indicate Euclidean balls $\{\mathcal{D}_x\}$ and plain segments represent the discrete balls $\{D_x\}$. (b) Pseudocode for the reduction process of $Sk$ points and illustration of the algorithm (arrows indicate the order in $\mathcal{L}$ or in $s$).

envelope. Since maximal elliptic paraboloids and maximal balls coincide, the points in $Sk$ belong to the medial axis. Some maximal elliptic paraboloids may not belong to $Sk$, as illustrated in Fig. 8 in the 1D case: The parabolas $\{A, B, C\}$ belong to the medial axis, whereas only the parabolas $A$ and $C$ belong to $Sk$ ($B$ is covered by the union of $A$ and $C$). To prove the second statement, we note that the definition of $Sk$ strictly coincides with the reverse distance transformation equations of Section 3. Once $Sk$ is computed, if we consider the strictly positive height values of the upper envelope elliptic paraboloids, we obtain the original shape. □

### 4.3 Reduced Discrete Medial Axis Extraction

In Section 4.2, we have proven that the skeleton $Sk$ is a subset of the medial axis in the continuous case. We first illustrate in this section that this property does not hold in the discrete case and present a process to transform $Sk$ points into maximal ball centers in the discrete case.

First, if we consider a binary shape in dimension $d$, we can easily transform the REDT algorithm to extract points in $Sk$: We first consider the image $F$ as the SDT of a binary object. Then, we apply the REDT algorithm in which we mark the upper envelope elliptic paraboloids to construct $Sk$. This labeling can easily be performed dimension after dimension since the array $s$ in Fig. 5 contains one-dimensional parabolas in the upper envelope.

In [33], Saito and Toriwaki use the $O(Avg.n^d)$ REDT process to extract the skeleton $Sk$. Using the optimal REDT algorithm proposed in the previous section, we obtain an algorithm to compute $Sk$ in $O(n^d)$, which is optimal for the problem.

In the 2D case, let us consider a binary shape and its skeleton $Sk$. We denote by $\{\mathcal{F}_x(i)\}_{i=0..N}$ the sequence of parabolas given by the intersection between the $Sk$ elliptic paraboloids and the column $j$ of the image. Hence, each parabola is such that $\mathcal{F}_x(i) = f(x, j) - (i - x)^2$. In this one-dimensional case, the differences between $Sk$ and the discrete medial axis (DMA for short) are illustrated in Fig. 9a: $\{D, E\}$ belong to $Sk$, whereas only $D$ belongs to the discrete medial axis. Another illustration of the differences between DMA and $Sk$ is presented in Fig. 10.

In the following, we detail a time optimal algorithm to extract a discrete medial axis from $Sk$.

We denote by $\mathcal{D}_x$ the disk associated with $\mathcal{F}_x(i)$ (i.e., a segment in the one-dimensional case). Furthermore, we consider the discrete disk $D_x$ associated with $\mathcal{D}_x$ as the set of discrete points contained in $\mathcal{D}_x$. To consider discrete maximal disk in $Sk$, we have to remove all points $x$ such that $D_x$ is not maximal. Given two parabolas of centers $x$ and $x'$, we have a simple test, denoted $\mathtt{Incl(x, x')}$, to decide if $D_x$ contains $D_{x'}$ (we just compare the ends of the segments). Let us denote by $[l_y, r_y]$ the interval given by a disk $D_y$. We consider the list $\mathcal{L}$ of parabolas sorted according to the left extremity of the segments. If some parabolas have the same left extremity coordinate, we sort such parabolas according to the right extremity position (see Fig. 9a, right). If $n$ denotes the size of the column $j$ in the image, the list $\mathcal{L}$ can be computed in $O(n)$ (we store the extremities in two arrays of size $n$ during the scan of the parabolas). If two segments are identical, we remove one of them and we label the other one with a flag

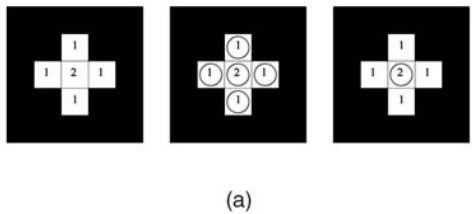

(a)                          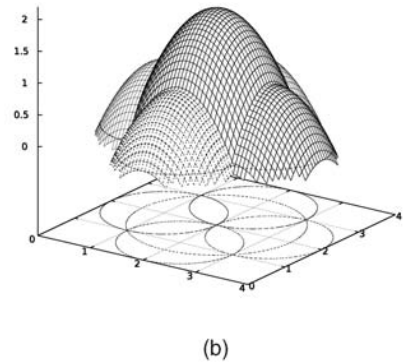

(b)

Fig. 10. (a) From left to right, a simple binary shape with its SDT, the points in $Sk$, and its discrete medial axis. (b) Illustration of the elliptic paraboloids in $Sk$.

"*double*" (see Definition 4). Using $\mathcal{L}$, we have a simple algorithm, presented in Fig. 9b, to remove all points that do not belong to the DMA from the set $\{\mathcal{F}_x(i)\}$. In this process, we scan the parabolas according to the $\mathcal{L}$ order and we test the inclusion of two consecutive parabolas in a greedy algorithm. Hence, the computational cost of this algorithm is $O(n)$ and the resulting set of parabolas is stored in the array $s$. The correctness of this algorithm is given by the following proposition:

**Proposition 2.** *The associated disk of a parabola is maximal if and only if the parabola belongs to $s$ at the end of the process.*

**Proof.** First of all, if the list $\mathcal{L}$ is reduced to one parabola, the associated disk is maximal and it belongs to $s$. We prove the proposition by induction. Let us consider the step $k$ ($k \geq 1$) in the algorithm in Fig. 9b. We suppose that, at this point, $s$ contains the maximal disk of the parabolas in $\{\mathcal{L}(i)\}_{0 \leq i \leq k}$ and we consider the disk $[u, v]$ of the parabola $\mathcal{L}(k+1)$. Note that the order of parabola in $s$ is the same as the order of parabola in $\mathcal{L}$. We denote by $[m, n]$ the segment associated with $s[q]$ (last inserted parabola in $s$). If the test $\text{Incl}(s[q], \mathcal{L}(k+1))$ is true, the segment $[m, n]$ contains the segment $[u, v]$ and, so, $\mathcal{L}(k+1)$ is not maximal and this parabola is not inserted in $s$. If we suppose that the inclusion test fails, $\mathcal{L}(k+1)$ is inserted in $s$. First of all, the segment $[u, v]$ cannot contain a segment in $s$. Indeed, by the definition of $\mathcal{L}$, if a parabola $x$ precedes the parabola $x'$ in $\mathcal{L}$, then the segment associated with $x'$ cannot contain the segment associated with $x$. Hence, $\mathcal{L}(k+1)$ does not change the maximal property of the segments in $s$. To complete the proof, we show that, if the test fails, no segment in $s$ contains the segment $[u, v]$. Let us consider a segment $[a, b]$ in $s$ such that $[a, b]$ contains $[u, v]$ and such that the segment $[a, b]$ is not associated with $s[q]$. So, we have $b \geq v$ and $a \leq u$. If the inclusion test fails between $[u, v]$ and $[m, n]$, then $v > n$ (we have $u \geq m$ by construction of $\mathcal{L}$). Hence, we have $b > v$. This leads to the contradiction that $[a, b]$ contains $[m, n]$ because the segments in $s$ are supposed to be maximal. Finally, it is sufficient to consider the inclusion test between $\mathcal{L}(k+1)$ and the last inserted parabola in $s$ to construct the set of maximal disks from the set $\{\mathcal{L}(i)\}_{0 \leq i \leq k+1}$. □

In higher dimensions, we apply this process to each dimension and we define the reduced discrete medial axis as follows:

**Definition 4 (Reduced Discrete Medial Axis).** *Let $P$ be a binary shape in dimension $d$ and $Q$ the SDT of $P$. We consider $Sk$ the Saito and Toriwaki's skeleton of $P$. The reduced medial axis (RDMA for short) is the set of points $(i, j)$ such that there exists at least one row in one of the $d$ dimensions in which the parabola associated with $(i, j, q(i, j))$ is preserved and not labeled "double" during the one-dimensional reduction process.*

**Theorem 1.** *Let $P$ be a binary shape in an image of dimension $d$ with $n^d$ grid points; the RDMA is a subset of the discrete medial axis of the shape, it has the reversibility property and the RDMA extraction is in $O(n^d)$.*

**Proof.** According to Corollary 1, $Sk$ is a subset of the continuous medial axis of the shape. Let us consider a discrete ball $B$, we prove that, if $B$ is preserved at the end of the reduction process, then $B$ belongs to the discrete medial axis. If we suppose that $B$ is not maximal in the discrete case, there exists another ball $B'$ such that $B'$ contains $B$. During the process, in each dimension, the segments associated with $B$ will either be removed or labeled "*double*" because they are contained in $B'$ segments. Hence, the ball $B$ will be removed from $Sk$. Finally, all resulting balls are maximal in the discrete plane. Furthermore, since the parabola removal process between two parabolas maintains the reversibility property, the final result allows us to reconstruct the shape. Concerning the computational cost, the $Sk$ computation is done in $O(n^d)$ and, for each row in each dimension, the one-dimensional step computational cost is linear in the number of parabolas in the row. Hence, the global cost of the reduction process is linear in the number of points in $P$, which is optimal for the problem. □

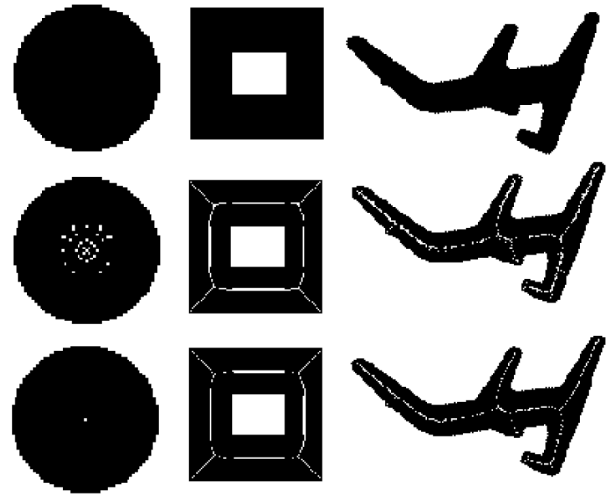

Fig. 11. Results of medial axis extraction in 2D: The first row presents the input binary shapes, the second row shows the $Sk$ sets (white pixels), and last the last row shows the RDMA points.

### 4.4 Results

For both the REDT and skeleton extraction algorithms in dimensions 2 and 3, Figs. 11 and 12 present results on several 2D and 3D shapes. In these figures, we compare the RDMA proposed in Section 4.3 to the $Sk$ set presented in Section 4.2.

The overall process can be sketched as follows: Given an input object, we first compute the SDT using the algorithm presented in Fig. 3. Then, we use the REDT algorithm of Fig. 5 and the parabola upper envelope computation process to construct the $Sk$ set. Finally, we use the RDMA extraction process presented in Section 4.3 to obtain the final set of balls. All of these algorithms have time optimal computational costs and C++ implementations are available.[1]

As expected, the RDMA contains fewer points and, thus, is a more compact reversible representation of the binary shapes. Table 1 gives details on the efficiency of the RDMA representation in the number of necessary information to have a lossless encoding of binary input shapes.

## 5 DISCUSSION

In this section, we link the REDT computation and RDMA extraction to classic tools in computational geometry and

---

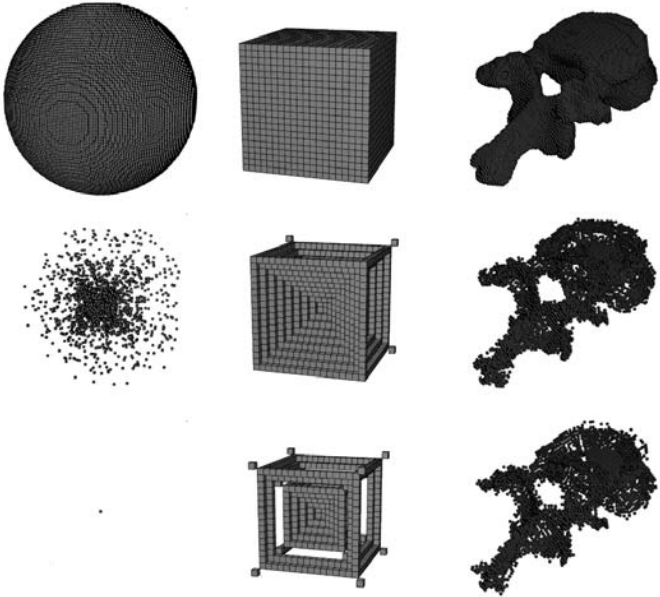

Fig. 12. Results of medial axis extraction on 3D objects: The first row presents the input binary shapes, the second one shows the $Sk$ sets, and the last one shows the RDMA points.

we will adapt some methods and objects from the computational geometry field for shape filtering.

### 5.1 Computational Geometry in the DT and MA Extraction Problems

First, we detail the links between the EDT and the construction of a Voronoi diagram [12], [14], [15].

Given a set of sites $\mathcal{S} = \{s_i\}$ in $\mathbb{R}^2$, the Voronoi diagram is a decomposition of the plane into cells $\mathcal{C} = \{c_i\}$ (one cell $c_i \subset \mathbb{R}^2$ per site $s_i$) such that, for each point $p$ in the (open) cell $c_i$, we have $d(p, s_i) < d(p, s_j)$ for $i \neq j$. In other words, $p$ is closer to the site $s_i$ than to any other site $s_j$ [39]. Let $\mathcal{V}_\mathcal{S}$ denotes the Voronoi diagram of $\mathcal{S}$. We define the *Voronoi labeling* denoted $V_\mathcal{S}$ as the intersection between $\mathbb{Z}^2$ and $\mathcal{V}_\mathcal{S}$. In other words, we assign to each grid point of the plane the index of the Voronoi cell containing it. We also consider the *closed cells* defined by $c_i^- = \{p \in \mathbb{R}^2; d(p, s_i) \leq d(p, s_j)\}$. The cell boundaries correspond to the loci of points equidistant to two or more sites. In the discrete plane, we can define a *Voronoi labeling* by assigning to each grid point the index of a Voronoi cell containing it. For grid points which are situated on the boundaries of the Voronoi diagram, the index of one of the neighboring cells is chosen arbitrarily.

If we now consider a binary object $P$ and its background $\bar{P}$, it is clear that, from the Voronoi labeling $V_{\bar{P}}$ of grid points $P$, we can extract the EDT of $P$. Indeed, if $p \in P$:

$$EDT(p) = d(p, V_{\bar{P}}(p)). \tag{15}$$

Hence, in the literature, we may find algorithms that compute the EDT based on a Voronoi labeling construction [12], [13], [15], [40]. We may also find algorithms that extract the Voronoi labeling from the EDT [16], [41].

In the following, we show that both the REDT computation and the RDMA extraction are based on a *Power diagram* (also known as the *Laguerre diagram*) construction [42]. First, we consider a set of sites $\mathcal{S} = \{s_i\}$ such that each point $s_i$ is associated with a radius $f(i)$. The power $\sigma_i(p)$ of a point $p$ in the plane according to the site $s_i$ is given by:

$$\sigma_i(p) = d(p, s_i) - f(i)^2. \tag{16}$$

If $\sigma_i(p) < 0$, $p$ belongs to the disk of center $s_i$ and radius $f_i$. If $\sigma_i(p) > 0$, $p$ is outside the disk. The *power diagram* is a kind of Voronoi diagram based on the metric induced by $\sigma$. Hence, the power diagram $\mathcal{V}'_\mathcal{S}$ is a decomposition of the plane into cells $C' = \{c'_i\}$ associated with each site $s_i$ such that:

$$c'_i = \{p \in \mathbb{R}^2 : \sigma_i(p) < \sigma_j(p), i \neq j\}. \tag{17}$$

Note that the cell $c'_i$ associated with a site $s_i$ may be empty; otherwise, $c'_i$ is a convex cell (see Fig. 13). The power diagram is a common tool in computational geometry when a geometry of spheres or hyperspheres must be taken into account [42], [43], [44].

### TABLE 1
Efficiency of the RDMA Representation of the 3D Binary Objects Presented in Fig. 12

| Object name | number of grid points | #Skel | #RDMA |
|---|---|---|---|
| Sphere of radius 50 | 523155 | 1767 | 1 |
| Cube of length 20 | 8000 | 940 | 624 |
| Vertebra | 103302 | 9264 | 6593 |

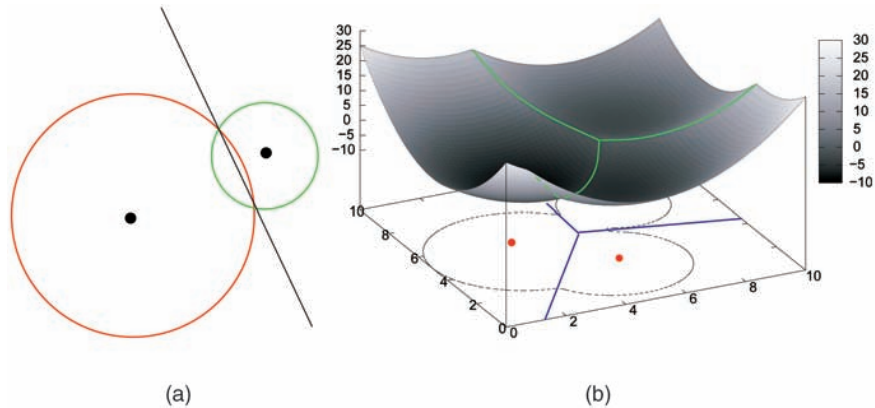

Fig. 13. (a) Power diagram of two sites with associated radii. (b) Illustration of the power diagram using elliptic paraboloids.

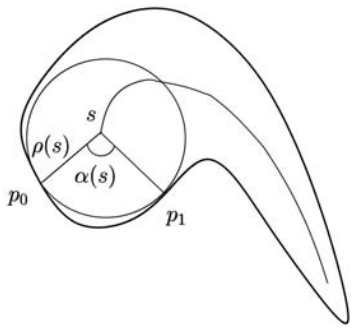

Fig. 14. The bisector angle at a point $s$ of the MA is defined by the angle $\widehat{p_0 s p_1}$, where $p_0$ and $p_1$ are the contact points between the maximal ball centered at $s$, with radius $\rho(s)$ and the shape boundary.

In the following, we define the *power labeling* $V'_S$ as the power diagram labeling of grid points. More precisely, we assign to each grid point of the plane the index of the cell containing it in the power diagram $\mathcal{V}'_S$. Let us consider a binary shape $P$ and its discrete medial axis (or RDMA) $L$ (see Section 3). Note that a radius is associated with each point in $L$.

**Proposition 3.** *Let $p$ be a discrete point and $i$ the index of the cell $V'_L(p)$, $p$ belongs to the shape $P$ iff:*

$$\sigma_i(p) < 0. \tag{18}$$

**Proof.** The proof is straightforward since (10) is the opposite of (16). Furthermore, (6) can be rewritten as:

$$P = \{p \in \mathbb{Z}^2 \mid \{-\sigma_i(p)\} > 0\}. \tag{19}$$

$\square$

Using Proposition 3, we can link the REDT to the computation of nonempty cells in the power labeling.

There is another strong analogy between the power diagram and RDMA extraction: As depicted in Fig. 13b, the power diagram can be constructed by the intersection between the plane $z = 0$ and the lower envelope of elliptic paraboloids defined by the power functions in (16). This illustration corresponds to the RDMA extraction algorithm in which we consider opposite elliptic paraboloids. Furthermore, if a disk $A$ (with center $p_a$ and radius $f(a)$) contains a disk $B$ (with center $p_b$ and radius $f(b)$), then the cell of $p_b$ is empty in the power diagram of points $p_a$ and $p_b$. Hence, we have an equivalence between maximal disks in the continuous plane and empty cells in the power diagram.

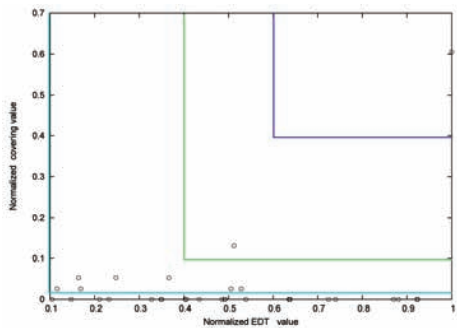
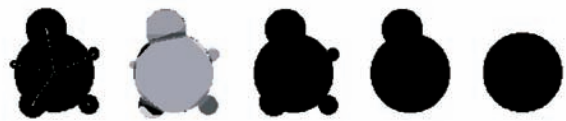

(a)

(b)

Fig. 15. Illustration of the filtering process in 2D: (a) The filtering graph (the x-axis represents the normalized $\rho$ values and the y-axis is the normalized $\kappa$ values are the covering and radius measures). (b) The original shape with its RDMA (38 points), the power labeling (pixels with same gray value belong to the same cell), and fine to coarse reconstructions (with 9, 2 points and 1 point).

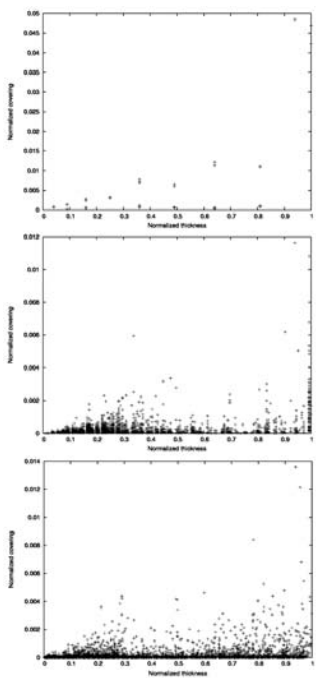
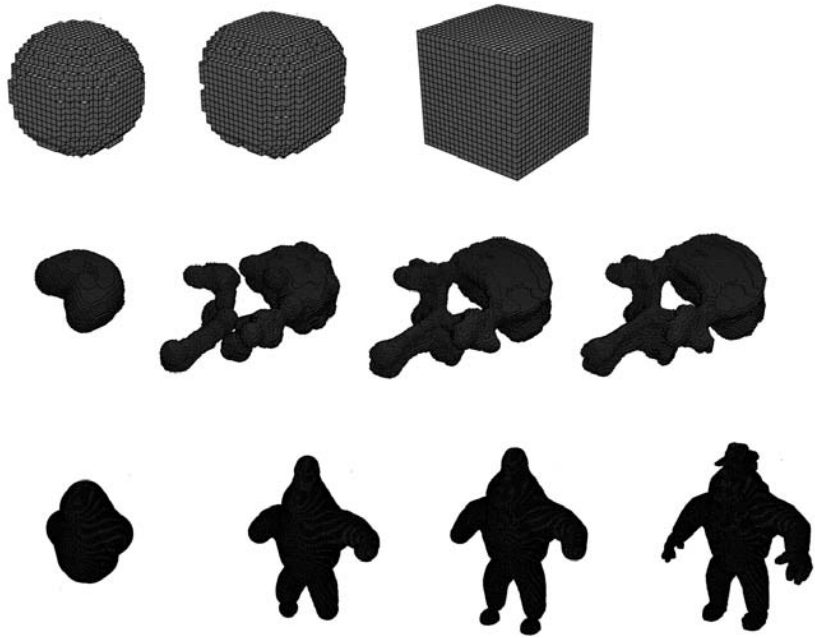

Fig. 16. Illustration of the filtering process in 3D: Distribution of RDMA points according to the values $\rho(p)$ and $\kappa(p)$ and coarse to fine reconstructions. The thresholds are given in Table 2.

Finally, if $P$ is a discrete shape where pixels are associated with their SDT and if $\mathcal{V}'_P$ is the power diagram of such points, then the set $Sk$ as defined in Section 4 is the set of sites in $V'_P$ such that $\mathcal{V}'_P \cap P \neq \emptyset$.

Finally, the RDMA corresponds to a subset of these sites according to the filtering procedure detailed in Section 4.3.

## 5.2 An Application to the Euclidean Discrete Medial Axis Filtering

In this section, we illustrate the use of the power diagram to process DMA points. In this case, we propose a different filtering process of MA points whatever the dimension is. In some applications, the MA or the DMA suffer from the presence of nonsignificant branches due to noises on the object boundary. If we consider nonreversible characterizations of the object, it may be interesting to filter MA balls up to a relevance criterion. Usually, the radius of the balls is one of these criterions. In dimensions 2 and 3, the bisector angle can also be used [37], [23] (see Fig. 14). However, generalizations to higher dimensions are not trivial. Note that the filtering processes discussed here do not provide a control on the global connectivity of the reconstructed shape.

In this paper, we propose a simple filtering process based on two criteria defined whatever the dimension. Given an object $P$ and for each point $p$ in the RDMA, we define two measurements:

- the *thickness*, $\rho(p)$, is the radius of the ball;
- the *covering*, $\kappa(p)$, corresponds to the number of grid points in the cell associated with $p$ in the power labeling of all RDMA points in $P$.

The thickness allows us to remove small balls, whereas the covering measurement $\kappa(p)$ represents the importance of the ball in the shape according to all other balls: If the area of the cell at $p$ is small, then the ball centered at $p$ is *covered* by the neighboring balls. If we remove $p$ from the RDMA, the difference in the number of grid points between the original object and the reconstructed one is bounded by $\kappa(p)$. This geometrical interpretation of the $\kappa(p)$ values is interesting to control the filtering quality.

Note that $\rho(p)$ can be normalized by the diameter of the shape and $\kappa(p)$ by the area of $P$.

The filtering process considers two thresholds, $\rho_0$ and $\kappa_0$, and a ball centered at $p$ belongs to the filtered medial axis if:

$$\rho(p) \geq \rho_0 \quad \text{and} \quad \kappa(p) \geq \kappa_0. \qquad (20)$$

Given a binary object in an image of dimension $d$ with $n^d$ grid, we have presented in Section 4.3 an $O(n^d)$ algorithm to extract the RDMA. Since the power labeling of RDMA points in $P$ corresponds to the extraction of the upper envelope of $d$-dimensional elliptic paraboloids, the measurements $\rho$ and $\kappa$ can be computed in linear time in the number of grid points.

Figs. 15 and 16 present some results of the filtering process in dimensions 2 and 3. In Fig. 15, the rectangular

### TABLE 2
Results in Number of RDMA Points of the Filtering Process of the Objects in Fig. 15a and in Fig. 16, right

| | $\rho_0$ | $\kappa_0$ | number of disks | |
|---|---|---|---|---|
| | 0 | 0 | 38 | |
| | 0 | 0.004 | 9 | |
| | 0.4 | 0.1 | 2 | |
| | 0.6 | 0.4 | 1 | |

| Object name | $\rho_0$ | $\kappa_0$ | number of balls | number of voxels |
|---|---|---|---|---|
| Cube of length 20 | 0 | 0 | 624 | 8000 |
| | 0.5 | 0.0005 | 24 | 6200 |
| | 0.5 | 0.025 | 8 | 5112 |
| Vertebra | 0 | 0 | 6593 | 103302 |
| | 0.1 | 0 | 5096 | 100055 |
| | 0.2 | 0.0005 | 304 | 78535 |
| | 0.5 | 0.001 | 75 | 41873 |
| "AI" 200x200x200 | 0 | 0 | 17098 | 549162 |
| | 0.1 | 0 | 8057 | 514068 |
| | 0.1 | 0.001 | 197 | 458541 |
| | 0.5 | 0.001 | 132 | 328823 |

Note that if thresholds are (0, 0), no RDMA points are removed.

areas in graphs contain the RDMA points that are removed during the filtering. The thresholds used and the obtained number of balls are presented in Table 2. On 3D examples, Table 2 also contains the size of the reconstructed shapes in the number of voxels.

## 6 CONCLUSION

In this paper, we first have optimized the REDT computation algorithm and have obtained a computation cost in $O(n^d)$ for a $d$-dimensional image which is time optimal ($n^d$ is the total number of grid points). Then, we have presented a $d$-dimensional reversible RDMA extraction algorithm in $O(n^d)$. We have proven that the proposed RDMA is a subset of the classic discrete medial axis of the shape. Beside these theoretical results, we have also detailed algorithms and provided a reference to C++ implementations. In the discussion, we have illustrated the strong links between the DT and the MA problems and classic problems in computational geometry. Based on these results, we have proposed a simple but efficient filtering process of $d$-dimensional RDMA using two parameters to control the shape of the reconstructed object.

In future works, we expect other optimizations of the RDMA extraction process to reduce the number of points. The final goal of this optimization should be to compute the *optimal* reversible skeleton of a shape (in the sense of having a minimal number of points, see [35], [36] for related papers).

Furthermore, the presented links between MA and computational geometry objects suggest many other developments to improve the description of a shape using different metrics or high-level features based on the union of maximal balls.

## ACKNOWLEDGMENTS

The authors would like to thank Dan Frost for his helpful comments on English language and the reviewers for their interesting comments.

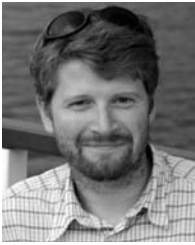

**David Coeurjolly** received the master's degree in computer science from the Ecole Normale Supérieure, Lyon, and the Université Claude Bernard, Lyon, France in 2000, and the PhD degree in computer science from the Université Lumière Lyon2, in 2002. He holds a permanent research position (Chargé de Recherche CNRS) at LIRIS Laboratory, Lyon. He chairs, with Annick Montanvert, Technical Committee 18 "Discrete Geometry" of the International Association for Pattern Recognition. His research interests include digital geometry, computational geometry, and image analysis.

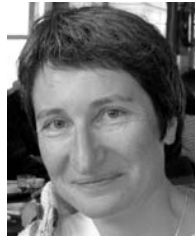

**Annick Montanvert** received the MS degree in computer science from the Ecole Nationale Supérieure d'Informatique et de Mathématiques Appliquées de Grenoble (ENSIMAG), Grenoble, France, and the PhD degree in computer science from the Joseph Fourier University, Grenoble, in 1984 and 1987, respectively. In 1989, she spent six months at the Center for Automation Research, University of Maryland. After being an assistant professor, she received the Habilitation á Diriger des Recherches in 1992 and she became a professor in computer science at the Ecole Normale Supérieure in Lyon. She is currently a professor at the Pierre Mendès France University, Grenoble, where she directed the Computer Science Department of the Technological Institute from 1996 to 2002. She performs her research at the Laboratory of Image and Signal (LIS), Grenoble. She is member of the steering committee of the DGCI Conference (Discrete Geometry for Computer Imagery) and she chairs together with David Coeurjolly, Technical Committee 18 (specialized in discrete geometry) of the IAPR (International Association for Pattern Recognition). Her research interests include discrete topology and geometry, shape representation, multiresolution segmentation and analysis, image analysis, image, and video processing.